\documentclass[aps,pre,reprint,amsmath,amssymb,superscriptaddress,showpacs,floatfix]{revtex4-1}
\usepackage{graphicx}
\usepackage{dcolumn}
\usepackage{bm}
\usepackage[colorlinks, allcolors=blue]{hyperref}
\usepackage[usenames, dvipsnames]{color}

\begin{document}

\title{Hybrid ED/DMRG approach to the thermodynamics of 1D quantum models}

\author{Sudip Kumar Saha}
\affiliation{S. N. Bose National Centre for Basic Sciences, Block - JD, Sector - III, Salt Lake, Kolkata - 700106, India}

\author{Dayasindhu Dey}
\affiliation{S. N. Bose National Centre for Basic Sciences, Block - JD, Sector - III, Salt Lake, Kolkata - 700106, India}

\author{Manoranjan Kumar}
\email{manoranjan.kumar@bose.res.in}
\affiliation{S. N. Bose National Centre for Basic Sciences, Block - JD, Sector - III, Salt Lake, Kolkata - 700106, India}

\author{Zolt\'an G. Soos}
\email{soos@princeton.edu}
\affiliation{Department of Chemistry, Princeton University, Princeton, New Jersey 08544, USA}

\date{\today}

\begin{abstract}

Exact diagonalization (ED) of small model systems gives the thermodynamics of spin chains or 
quantum cell models at high temperature $T$. Density matrix renormalization group (DMRG) calculations of progressively 
larger systems are used to obtain excitations up to a cutoff $W_C$ and the low-$T$ thermodynamics. The hybrid approach 
is applied to the magnetic susceptibility $\chi(T)$ and specific heat $C(T)$ of spin-$1/2$ chains with isotropic exchange 
such as the linear Heisenberg antiferromagnet (HAF) and the frustrated $J_1-J_2$ model with ferromagnetic (F) $J_1 < 0$ and 
antiferromagnetic (AF) $J_2 > 0$.  The hybrid approach is fully validated by comparison with HAF results. It extends $J_1-J_2$ 
thermodynamics down to $T \sim 0.01|J_1|$ for $J_2/|J_1| \geq \alpha_c = 1/4$ and is consistent with other methods. The criterion 
for the cutoff $W_C(N)$ in systems of $N$ spins is discussed. The cutoff leads to bounds for the thermodynamic limit 
that are best satisfied at a specific $T(N)$ at system size $N$.
	
\end{abstract}

\maketitle

\section{\label{sec:intro}Introduction}
White introduced the density matrix renormalization group (DMRG) and applied it to the ground state 
of quantum spin chains~\cite{white-prl92, *white-prb93}. DMRG has become a powerful general method for 
the ground state and excitation gaps that characterize the quantum ($T = 0$) phases of one-dimensional 
(1D) models with spin or charge degrees of freedom~\cite{schollwock2005, karen2006}. The transfer matrix 
renormalization group (TMRG) is a related approach to finite temperature in which the partition function 
with increasing system size is followed to lower $T$~\cite{nishino1995,peschel99,huang2012}. White and Feiguin generalized DMRG 
to finite $T$ by enlarging the Hilbert space~\cite{feiguinprl2004, *feiguinprb2005}. The auxiliary Hamiltonian 
contains fictitious states in one-to-one correspondence with the physical basis. Karrasch et al. discuss ways 
to facilitate the calculation of transport properties using the time dependent DMRG at finite $T$~\cite{karrasch2012, *karrasch2013}. 
These methods have strengths and limitations. DMRG has been applied directly to the low-$T$ thermodynamics of gapped chains with two spins 
per unit cell~\cite{pati1997}. The 
striking success of DMRG at $T = 0$ provides strong incentive for extension to finite $T$. The most challenging systems are 
gapless chains with one spin per unit cell and quasi-long-range correlations in the ground state.  

We develop in this paper a hybrid approach to the thermodynamics of spin chains and quantum cell models. 
The high-$T$ regime is treated conventionally by exact diagonalization (ED) of small systems. DMRG then gives 
the low-energy excitations of increasingly large systems. Partition functions based on a few thousand states 
yield the low-$T$ thermodynamics. The combination of ED and DMRG covers the entire range, down to $T$ set by the 
accuracy of DMRG excitations, without ever invoking the full spectrum of large systems. The hybrid ED/DMRG approach 
is general, with DMRG tuned to the low-energy spectrum instead of the ground state.

There are broadly two contexts, mathematical and physical, for discussing spin chains or 1D quantum cell models. 
The spin-$1/2$ linear Heisenberg antiferromagnet (HAF) is the oldest and best characterized many-spin system~\cite{bethe31, 
*hulthen38, johnston2000}. The spin-$1$ HAF or other spin-$1/2$ models have been intensively studied for decades using field 
theory~\cite{chubukov1991, hikihara2008} and numerical methods~\cite{sandvik2010}. Correlated many-spin or many-electron models 
are intrinsically interesting. The characterization of quasi-1D compounds with linear chains of transition metal ions or 
organic radical ions has an equally long history~\cite{jongh1974,miller83}. Isotropic exchange is the dominant interaction, but not 
the only one. Thermodynamics to a factor of two or three lower $T$ than possible by ED would significantly aid the analysis 
of magnetic data. The $T \rightarrow 0$ limit is interesting mathematically.   

The $J_1-J_2$ model, Eq.~\ref{eq:j1j2} below, illustrates both contexts. The model has one spin-$1/2$ per unit cell and isotropic exchange 
$J_1$ and $J_2 > 0$ with first and second neighbors, respectively. The quantum phases for AF exchange $J_1 > 0$ include the 
exact ground state at the Majumdar-Ghosh~\cite{ckm69b} point ($J_2 = J_1/2$) and the critical point~\cite{nomura1992} $J_2/J_1 = 0.2411$ 
at the onset of the gapped dimer phase. The quantum phases for F exchange $J_1 < 0$ feature the critical point~\cite{hamada88} at 
$J_2 = |J_1|/4$ between the FM ground state and the gapped incommensurate (IC) singlet ground state~\cite{sirker2011}. The gapless 
decoupled phase~\cite{soos-jpcm-2016, *soos2013} includes $J_1 = 0$ and lies between IC phases with $J_1 < 0$ and $J_1 > 0$. 

The $J_1-J_2$ model with $J_1 < 0$ is the starting point for the magnetic properties of cupric oxides that contain chains of 
spin-$1/2$ Cu(II) ions and have singlet ground states~\cite{hase2004, drechsler2007, dutton2012, masuda2004, *park2007, wolter2012}. 
An applied magnetic field of $10$ Tesla is sufficient to induce the FM ground state in some cases. The $T$ and field 
dependencies of the magnetization and magnetic specific heat can be followed in systems with competing F and AF interactions. 
Present estimates of $J_1$ and $J_2$ in specific materials are rather approximate. At issue are the low-$T$ thermodynamics 
of the model, corrections due to spin-orbit coupling and additional (dipolar, hyperfine, interchain) weak interactions. 
We discuss the zero-field thermodynamics and focus on the magnetic susceptibility and specific heat.

The hybrid ED/DMRG approach is applicable to quantum cell models with a large but finite basis that increases exponentially 
with system size. There are $(2S + 1)^N$ states in a system of $N$ spins-$S$, and similar expressions hold for models with 
charge as well as spin degrees of freedom. Here we consider N spins-$1/2$ in models indexed by $\alpha$. The energy spectrum 
$\lbrace E(\alpha,N) \rbrace$ has $2^N$ states for any $\alpha$. The thermodynamics is governed by the canonical partitions function
\begin{equation}
Q(T,\alpha,N) = \sum_j \exp \left( -\beta E_j(\alpha,N) \right),
\label{eq:full_partition}
\end{equation}
where $T$ is the absolute temperature, $\beta=1/k_B T$, $k_B$ is the Boltzmann constant, the sum is over all states, and $E_j(\alpha,N)$ 
is relative to the ground state energy. The per spin result of the infinite chain is
\begin{equation}
N^{-1} \ln{Q(T,\alpha,N)} \rightarrow \ln{Q(T,\alpha)}.
\label{eq:part_thermolimit}
\end{equation}
The problem is to obtain or approximate the thermodynamic limit.

Our basic premise is that the full spectrum $\lbrace E(\alpha,N) \rbrace$ of large systems is never needed. The most demanding 
cases are gapless chains with quasi-long-range order in the ground state or chains with exponentially small gaps. 
Even then, thermal fluctuations suppress correlations between distant spins and the system size becomes irrelevant 
when $N$ is several times the correlation length. ED yields the full spectrum $\lbrace E(\alpha,N) \rbrace$ up to $N$, here to 
$N = 24$ for spin-$1/2$ chains. We can always find $T_n(\alpha,N)$ such that the thermodynamic limit is satisfied 
at $T > T_n(\alpha,24)$ for the quantity of interest. The low energy part of $\lbrace E(\alpha,N) \rbrace$ for larger $N$ is required 
at lower $T$, and DMRG is well suited for low-energy excitations. In principle, the problems are to obtain the low-energy 
excitations and to combine them with ED results.

\begin{figure}
\includegraphics[width=\columnwidth]{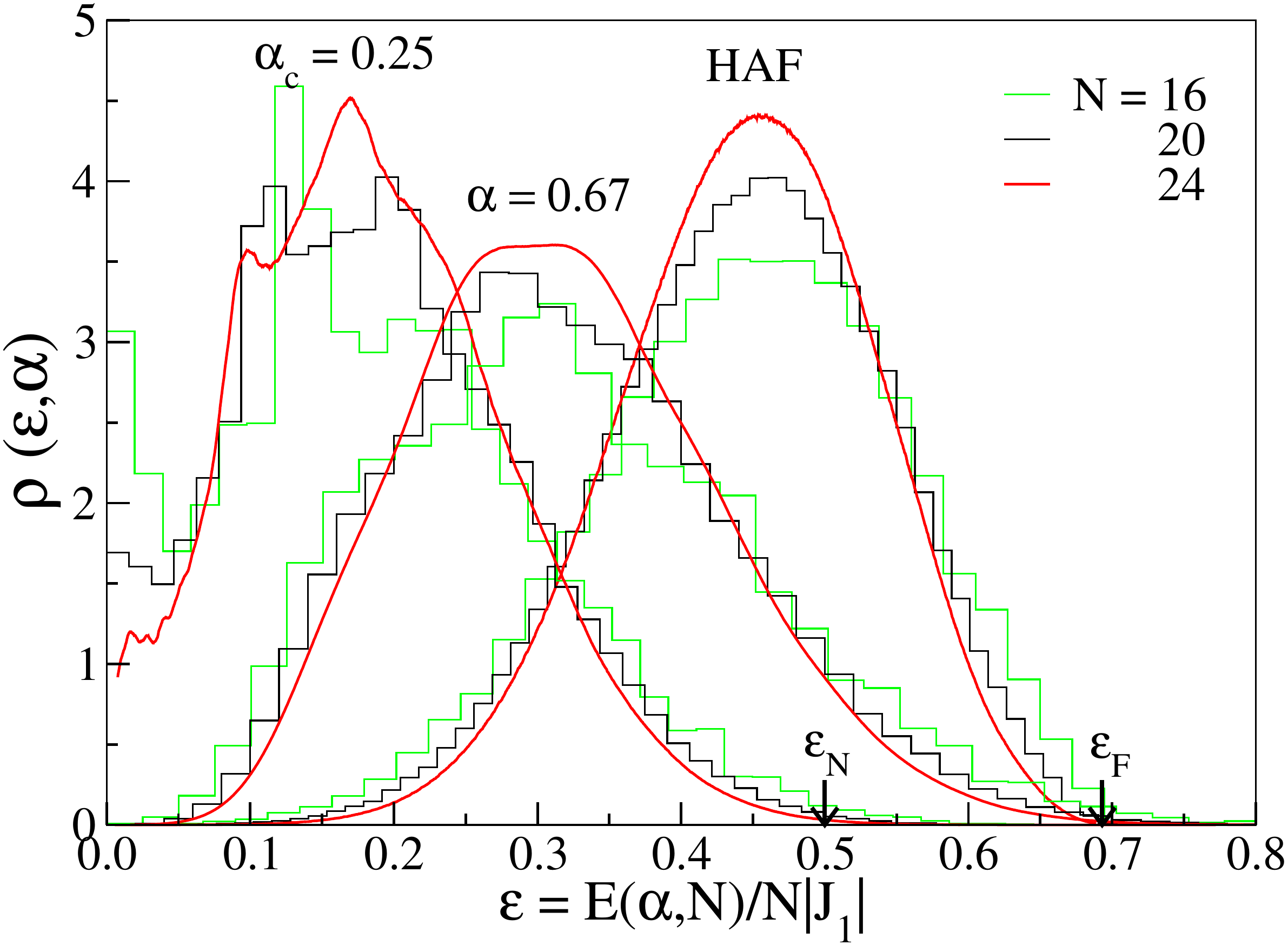}
\caption{\label{fig1} Normalized density of excitations, $\rho (\varepsilon,\alpha)$ with 
$\varepsilon = E(\alpha,N)/N|J_1|$, for three models of $N$ spins. The ground state is a singlet 
$(S = 0)$ at $\varepsilon=0$. The N\'eel state $(\ldots \alpha \beta \alpha \beta \ldots)$ has excitation 
energy $\varepsilon_N = 1/2$ for the $\alpha_c$ model while the FM state 
$(\ldots \alpha \alpha \alpha \alpha \ldots)$ has $\varepsilon_F=\ln2$ for the HAF.}
\end{figure}

The same conclusion follows from the increasing density of states with system size and the passage from a sum in 
$Q(T,\alpha,N)$ to an integral over excitations. The Boltzmann factor varies smoothly at high $T$ but is strongly peaked 
at low $T$. We expect and find lower $T_n(\alpha,N)$ in models with a high density of low-energy excitations. The 
normalized density of excitations, $\rho (\varepsilon,\alpha)$ with $\varepsilon = E(\alpha,N)/N|J_1|$, is shown in Fig.~\ref{fig1}
for $N = 16$, $20$ and $24$ spins in the HAF and $J_1-J_2$ models with $J_1 < 0$ and $\alpha= J_2/|J_1|$. We obtained $\rho (\varepsilon,\alpha)$ 
for $N = 16$ and $20$ as the number of states in $40-50$ bins of equal width. We took narrower bins for $N = 24$ and 
averaged over several adjacent bins to get relatively smooth curves.

A triplet at $E_T(\alpha,N)$ is typically the lowest excitation of spin chains with a singlet ground state. 
The size dependence of $E_T(\alpha,N)$ has been extensively discussed for the HAF and half-filled Hubbard or 
extended Hubbard models. The spin chains we consider have frustrated exchange interactions that shift the density 
of excitations to lower energy in Fig.~\ref{fig1}. We shall be comparing systems with similar finite-size gaps but 
different thermodynamics that reflects the excitation spectrum $\rho (\varepsilon,\alpha)$.

The paper is organized as follows. Section~\ref{sec2} presents the hybrid ED/DMRG procedure, starting with 
the DMRG calculation of excitation energies, moving on to truncation and ending with comparison with HAF thermodynamics. 
We turn in Section~\ref{sec3} to the $J_1-J_2$ model with F exchange $J_1 < 0$ and AF exchange $J_2 > 0$.  We compare results 
on the $J_1 < 0$ side with field theory and TMRG. ED to $N = 24$ accounts well for the spin susceptibility $\chi(T)$ and specific 
heat $C(T)$ per site at the critical point, $\alpha_c = 1/4$. DMRG extends the thermodynamics of F/AF chains to $T \sim 0.01$ in 
units of $|J_1|$, the largest exchange. Section~\ref{sec4} is a brief discussion of the method and its scope.  
\section{\label{sec2} Thermodynamics, Truncation and Extrapolation}
We develop in this Section the thermodynamics of spin chains without invoking the full energy spectrum 
$\lbrace E(\alpha,N) \rbrace$. In~\ref{Sec:2A} we obtain the low-energy states $E_j(\alpha,N)$ of models $\alpha$ with $N$ 
spins. In~\ref{Sec:2B} we truncate the partition function $Q(T,\alpha,N)$ in Eq.~\ref{eq:full_partition} at 
$E_j(\alpha,N) \leq W_C(\alpha,N)$ and discuss the choice of the energy cutoff. Extrapolation to the thermodynamic 
limit is demonstrated in~\ref{Sec:2C}  against exact HAF results. The applications in Section~\ref{sec3} are to $J_1-J_2$ models for 
which numerical analysis is more difficult and exact results are limited to $T = 0$.

The $J_1-J_2$ model has isotropic exchange $J_1$, $J_2/|J_1| = \alpha$ and is frustrated for either sign of $J_1$. We consider $J_1 < 0$ and set $|J_1| = 1$ 
as the unit of energy 
in chains of $N = 4n$ spins-$1/2$ with periodic boundary conditions. The model Hamiltonians $\alpha$ are
\begin{equation}
H(\alpha) = -\sum_{r} \vec{S}_r \cdot \vec{S}_{r+1} + \alpha \sum_{r} \vec{S}_{r} \cdot \vec{S}_{r+2}.
\label{eq:j1j2}
\end{equation}
The ground state is a singlet, total $S = 0$, for $\alpha > \alpha_c = 1/4$. The singlet and FM states 
are degenerate at the exact quantum critical point $\alpha_c$ ~\cite{hamada88}. The degeneracy at $\alpha_c$ 
is also exact for finite $N = 4n$ ~\cite{mk2012}. The HAF has AF exchange $J_1 = 1$ and $\alpha = 0$ in Eq.~\ref{eq:j1j2}. 
Exact, field theoretical and numerical results for its thermodynamics in zero field are summarized in detail in Ref.~\onlinecite{johnston2000}.  
Although there are open questions, especially in finite field, nowadays the HAF provides convenient tests of numerical methods. 

\subsection{\label{Sec:2A}DMRG}
We use the efficient DMRG algorithm for periodic boundary conditions in Ref.~\onlinecite{ddpbc2016}, where it was applied to the ground 
state energy and lowest excitation of HAFs with spin-$1/2$ and $1$. The superblock in this method has two new sites 
in addition to the left and right blocks. Since Eq.~\ref{eq:j1j2} has second neighbor interactions, we take new blocks 
of two sites in order to avoid interaction terms between old blocks. Four sites are added in each block at every step 
of infinite DMRG. The accuracy and computational costs are similar to matrix product state calculations~\cite{ddpbc2016}.

Infinite DMRG is used to generate the desired system of $N = 4n$ spins. Some 5-10 sweeps of finite DMRG are then performed. 
In most calculations we kept $m = 400$ eigenvectors that correspond to highest eigenvalues of the system block density matrix. 
The superblock Hamiltonian has dimension $m^2 2^4$. The ground state $E_1(N)$ is taken as zero. The states $j > 1$ have excitation 
energies $E_j(N) > 0$. The DMRG partition function with $l$ states of the superblock Hamiltonian is
\begin{equation}
Q_l(T,N) = \sum_{j=1}^l \exp \left( -\beta E_j(N) \right).
\label{eq:part_dmrg}
\end{equation}
We later consider truncated partition functions $Q_C(T,N)$ with $E_j(N) \leq W_C(N)$ at energy cutoff $W_C(N)$. 

We introduce in this paper several modifications that are tailored for finite systems. The focus is on 
excitations rather than the ground state. To improve the accuracy of the spectrum, we construct the system 
block density matrices $\rho_j(N)$ for the $l$ levels at system size $N$ and define an effective density matrix $\rho^\prime(\beta^\prime,l)$
\begin{equation}
\rho^\prime(\beta^\prime,l,N) = \sum_{j=1}^l \rho_j(N) \exp \left( -\beta^\prime E_j(N) \right)/Q_l(T,N).
\label{eq:denmat_trunc}
\end{equation}
The $l = 1$ case is simply $\rho^\prime(\beta^\prime,1)=\rho_1$ when the ground state is sought. Contributions 
for $l > 1$ are governed by $\beta^\prime$, an effective inverse $T$. We set $ \beta^\prime = 10$ (in units of $1/|J_1|$) 
since $T \sim 0.1$ is the range of interest. Variations of $\beta^\prime$ by 10 to 20$\%$ hardly change the accuracy of
the spectrum. The effective density matrix becomes important when the lowest excitations are closely spaced.

The system block Hamiltonian and all operators are renormalized by $\rho^\prime(\beta^\prime,l,N)$ to obtain the 
energy spectrum the model Hamiltonian at system size $N$. We perform two calculations. We first take $l = 5$ or $10$ 
in order to obtain the lowest excitations very accurately. The second calculation has $l > 100$. The entire spectrum 
is red shifted by an approximately constant amount because the density matrix now has projections from many excited 
states. Accordingly, we shift the spectrum by a constant and use the first calculation for the lowest excitations.

\begin{table}
\caption{\label{tab1}Exact (ED) and DMRG excitations in units of $|J_1|$ of chains
with $N = 24$ and 32 at $\alpha = 2/3$ in Eq.~\ref{eq:j1j2}. The ground state is at zero energy.}
\begin{ruledtabular}
\begin{tabular}{ c  c  c  c  c }
$\alpha = 2/3$ & \multicolumn{2}{c}{$ N = 24$} & \multicolumn{2}{c}{$ N = 32$} \\ \hline
 Level no. &  E (ED) & E (DMRG) & E (ED) & E (DMRG)  \\ \hline
2    &      0.1936     &        0.1936   &      0.1273   &      0.1283     \\
3    &      0.1936     &        0.1936   &      0.1397   &      0.1403     \\
4    &      0.2168     &        0.2169   &      0.1397   &      0.1405     \\
5    &      0.2299     &        0.2301   &      0.1541   &      0.1553     \\
6    &      0.2417     &        0.2418   &      0.1643   &      0.1659     \\
7    &      0.2701     &        0.2703   &      0.1866   &      0.1879     \\
8    &      0.2701     &        0.2703   &      0.1866   &      0.1883     \\
9    &      0.2817     &        0.2818   &      0.1883   &      0.1903     \\
10   &      0.2817     &        0.2821   &      0.1883   &      0.1907     \\
\end{tabular}
\end{ruledtabular}
\end{table}
\begin{table}
\caption{\label{tab2}Exact (ED) and DMRG excitation energies
for $N = 24$ and 32 at $\alpha = 1/2$ in Eq.~\ref{eq:j1j2}. The starred excitation
is the lowest singlet, $S=0$.}
\begin{ruledtabular}
\begin{tabular}{ c  c  c  c  c }
$\alpha = 1/2$ & \multicolumn{2}{c}{$ N = 24$} & \multicolumn{2}{c}{$ N = 32$} \\ \hline
 Level no. &  E (ED) & E (DMRG) & E (ED) & E (DMRG)  \\ \hline
2       &  0.0114*    &    0.0114*     &   0.0247*    &    0.0251*   \\
3       &  0.0522     &    0.0522      &   0.0385     &    0.0391    \\
4       &  0.0623     &    0.0624      &   0.0413     &    0.0418    \\
5       &  0.0623     &    0.0624      &   0.0413     &    0.0419    \\
6       &  0.0948     &    0.0949      &   0.0864     &    0.0876    \\
7       &  0.1144     &    0.1145      &   0.0947     &    0.0959    \\
8       &  0.1144     &    0.1145      &   0.0947     &    0.0961    \\
9       &  0.1256     &    0.1256      &   0.1027     &    0.1039    \\
10      &  0.1256     &    0.1258      &   0.1027     &    0.1054    \\
\end{tabular}
\end{ruledtabular}
\end{table}

To illustrate the accuracy, we compare DMRG excitation energies for $l = 400$ and $\beta^\prime = 10$ with exact results. 
The lowest 10 levels are listed in Table~\ref{tab1} for $N = 24$ and 32 at $\alpha= 2/3$, and in Table~\ref{tab2} at $\alpha= 1/2$. The $\alpha= 1/2$ 
levels are clearly denser than the $\alpha=2/3$ levels that in turn are denser than the corresponding HAF levels (not listed). 
Translational symmetry for periodic boundary conditions makes possible the ED results in the Tables. The accuracy of the lowest 
5 excitations is about 1 and 1.5$\%$, respectively, for $\alpha= 2/3$ and $1/2$. The HAF accuracy is better than 1$\%$. The accuracy 
up to level 100 is better than 5$\%$ and better than 10$\%$ for levels far higher than 100. Truncated partition functions are limited to $T \sim T^\prime(N)$ 
that depends on system size as discussed below. Since the cutoff $ W_C(N)$ is more than $ 10 T^\prime(N)$, the Boltzmann factors are very small. Accurate excitation 
energies are essential at low $T$.

To summarize, DMRG yields the excitations $E_j(\alpha,N) \leq W_C(\alpha,N)$ in models $\alpha$ with $N$ spins in Eq.~\ref{eq:j1j2}. Calculations 
are performed in sectors with Zeeman component $S^z = M$. The absolute ground state is in the $M = 0$ sector for $\alpha >\alpha_c$ and 
the $E_j(M,\alpha,N)$ are relative to $E_1(\alpha,N) = 0$.

\subsection{\label{Sec:2B}Thermodynamics}
The evolution of any thermodynamic quantity can be followed as the cutoff $W_C(N)$ is increased. The truncated partition function 
$Q_C(T,N)$ with $E_j(N) \leq W_C(N)$ in Eq.~\ref{eq:part_dmrg} is accurate at low $T$ and merges with ED at $T > T_n(N)$ when 
the full spectrum is retained. However, computational resources limit $W_C(N)$ and thermodynamics to $T < T_n(N)$, as seen explicitly 
for ED at $N = 24$. We need a criterion for choosing the cutoff. $W_C(N)$ leads to $R_C(M,N)$ states in sectors with $S^z = M$. 
Since $S$ is conserved, the total number of states is
\begin{equation}
R_C(N) = R_C(0,N)+ \sum_{M=1}^{N/2} 2R_C(M,N).
\label{eq:num_state}
\end{equation}
The number of states in the $M = 0$ sector is more convenient and intuitive than $W_C(N)$ for discussing thermodynamics. 
We retain $10^3-10^4$ states at low $T$ out of $2^N$ states. 

We chose $W_C(N)$ based on the maxima of $S_C(T,N)/T$ and $\chi_C(T,N)$, where $S_C(T,N)$ is the entropy per spin and $\chi_C(T,N)$ 
is the magnetic susceptibility per spin. Both are reduced at low $T$ by finite size gaps and at high $T$ by truncation.

\begin{figure}
\includegraphics[width=\columnwidth]{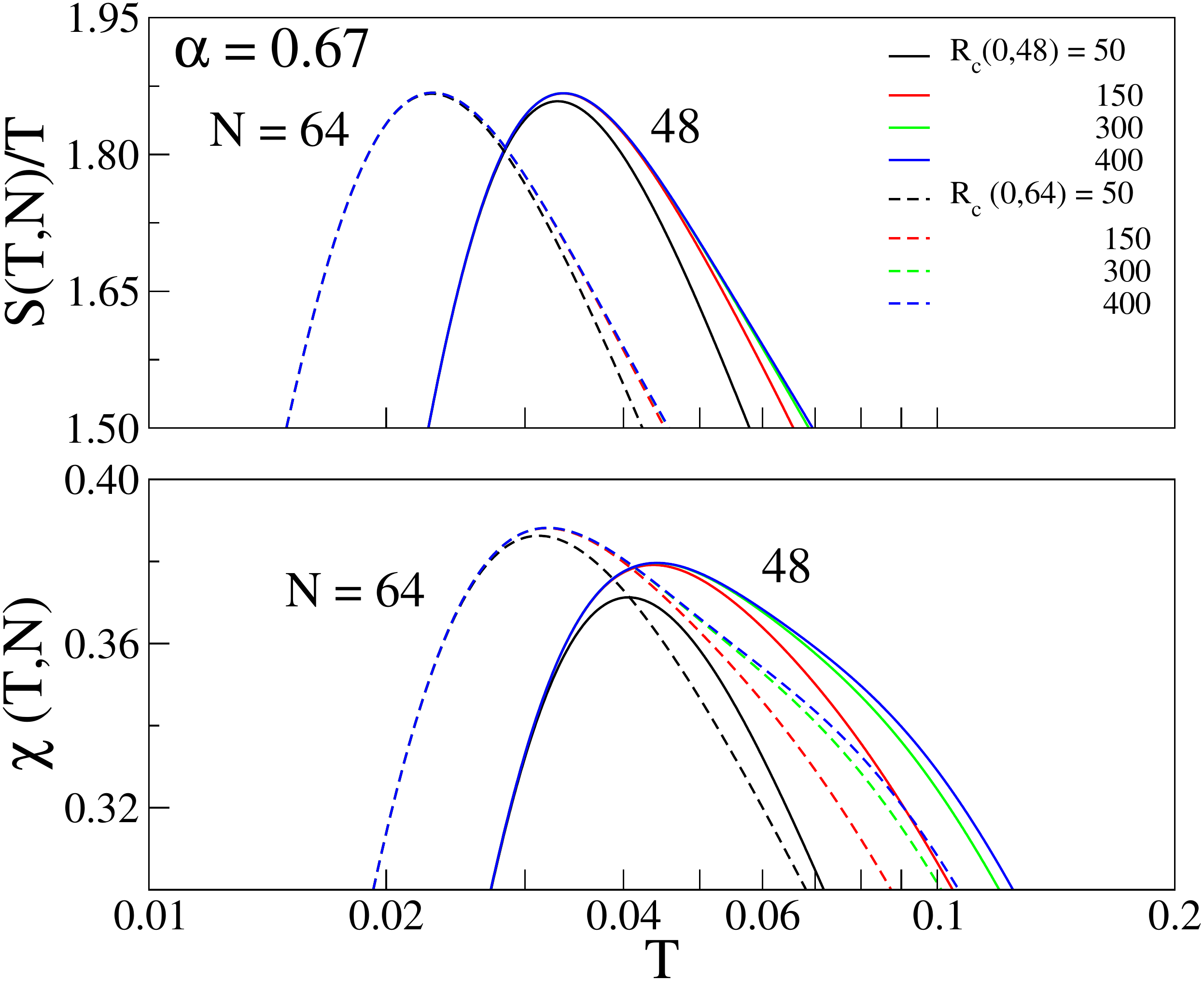}
\caption{\label{fig2} Convergence of the truncated $S_C(T,N)/T$ and $\chi_C(T,N)$ with increasing number of states in the $M = 0$ sector for $N = 48$ and 64 
spins at $\alpha=2/3$ in Eq.~\ref{eq:j1j2}. The maxima are converged when the cutoff $W_C(N)$ leads to $R_C(0,N) = 400$ states.}
\end{figure}

Fig.~\ref{fig2} 
illustrates the convergence of $S_C(T,N)/T$ and $\chi_C(T,N)$ for $N = 48$ and 64 at $\alpha= 2/3$ in Eq.~\ref{eq:j1j2}. 
The logarithmic scale is to emphasize low $T$. The cutoff governs the number of states in the $M = 0$ sector. $R_C(0,N) = 400$ ensures adequate 
convergence with respect to finite size gaps. The truncated partition function has $R_C(N) = 1563$ and 1818 states, respectively, at $N = 48$ and 64.

\begin{figure}
\includegraphics[width=\columnwidth]{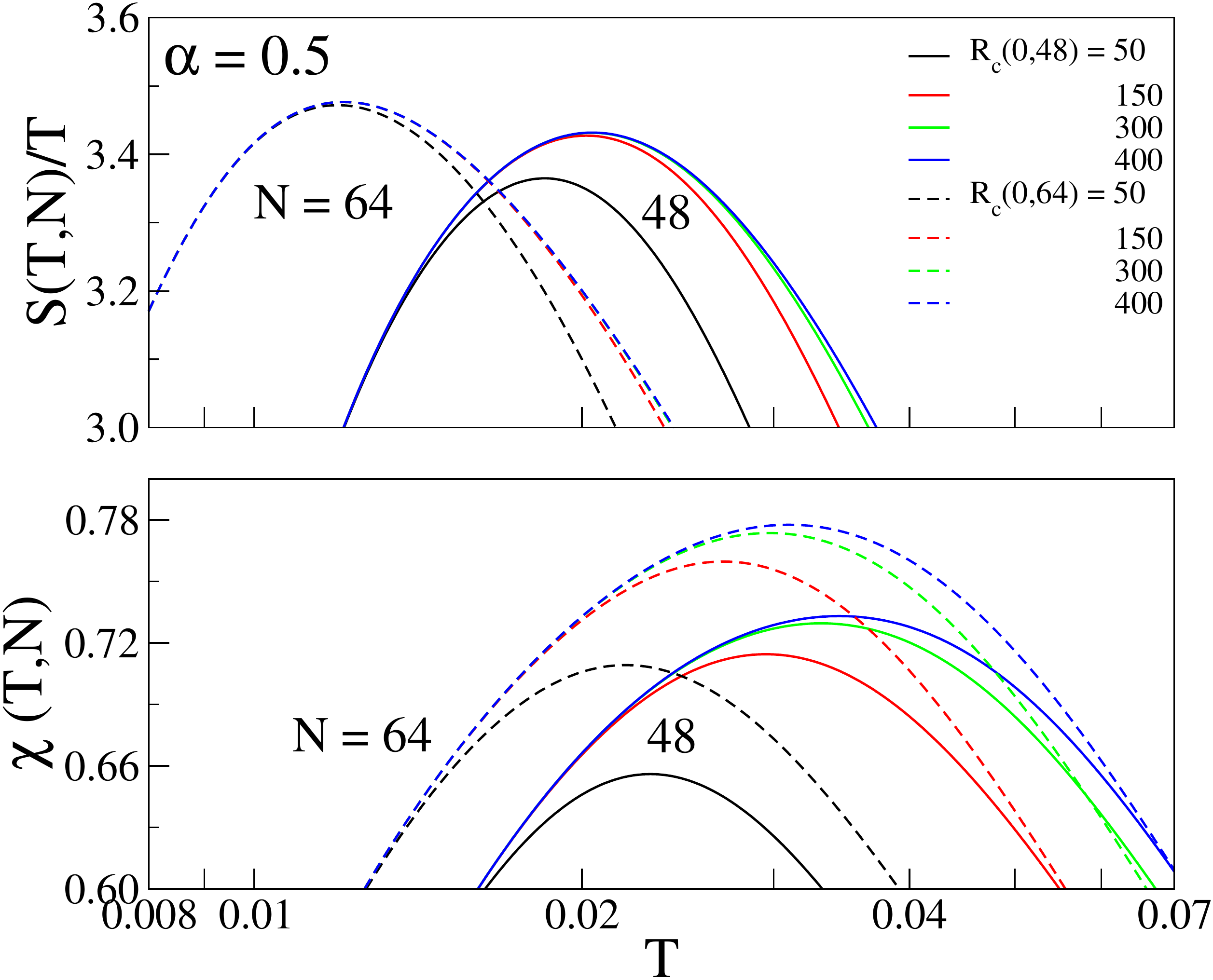}
\caption{\label{fig3} Convergence of the truncated $S_C(T,N)/T$ and $\chi_C(T,N)$ with increasing number of states in the 
$S^z = 0$ sector for $N = 48$ and 64 spins at $\alpha=1/2$ in Eq.~\ref{eq:j1j2}. The maxima are converged for 
cutoff $W_C(N)$ leads to $R_C(0,N) = 400$ states.}
\end{figure}

Fig.~\ref{fig3} shows the same functions for $\alpha=1/2$ in Eq.~\ref{eq:j1j2}. The maxima of $S_C(T,N)/T$ and $\chi_C(T,N)$ are 
about twice as high and are shifted to lower $T$ compared to $\alpha= 2/3$. However, the maxima are again converged with $R_C(0,N) = 400$ 
states. Now the truncated partition function has $R_C(N) = 1832$ and 2200 states, respectively, at $N = 48$ and 64. As implied by the $S(T)/T$ 
panel, there are many states with $E_j(64) < 0.01$ where the numerical accuracy has to be considered. The $\alpha= 1/3$ spectrum has even 
smaller and denser excitations. 

The full and truncated partition functions are given in Eq.~\ref{eq:full_partition} and Eq.~\ref{eq:part_dmrg}. 
Truncation always reduces $Q(T,N)$. It also reduces the internal energy $E(T,N)$ as shown by taking the 
difference and noting that the sum below is over $E_j(N) > W_C(N) > E_C(T,N)$,
\begin{equation}
\begin{aligned}
 & E(T,N)-E_C(T,N) = \frac{1}{Q(T,N)} \times \\
& \qquad \sum_j \left( E_j(N)-E_C(T,N) \right) \exp \left( -\beta E_j(N) \right). 
\label{eq:internal_ene}
\end{aligned}
\end{equation}
It follows that truncation also reduces the entropy $S(T,N) = k_B \ln Q(T,N)/N + E(T,N)/NT$. Truncation is 
arbitrarily accurate for $\beta W_C(N) \gg 1$ and inevitably fails at high $T$.

We will necessarily be 
working with $S_C(T,N)$ in large systems. The function $S_C(T,N)/T$ has a maximum at $T^\prime(N)$ where
\begin{equation}
S_C(T^\prime,N)=T^\prime(N)S^\prime_C(T^\prime,N).
\label{eq:entropy_specific}
\end{equation}
The same relation holds for the maximum of $S(T,N)/T$ or of $S(T)/T$. The maxima at $S_C(T^\prime,N)/T^\prime$ in 
Fig.~\ref{fig2} and Fig.~\ref{fig3} are lower bounds on $S(T)/T$ in the thermodynamic limit. They are the most 
accurate approximation at truncation $W_C(N)$. Accordingly, the cutoff criterion is convergence at the maximum.

Truncation reduces the entropy but not necessarily the susceptibility. The difference between the full and 
truncated magnetic susceptibility per site is

\begin{equation}
\begin{aligned}
& \chi(T,N)-\chi_C(T,N)= \frac{1}{NTQ(T,N)} \times \\
& \qquad \sum_j \left(M_j^2(N)-M_C^2(T,N) \right) 
\exp \left( -\beta E_j(N) \right).
\label{eq:sus_diff}
\end{aligned}
\end{equation}
The sum is over states $E_j(N) > W_C(N)$ with Zeeman components $S^z = M_k$, and $M_C^2(T,N)$ is the average value 
of $M^2$ over $E_j(N) \leq W_C(N)$. There is no guarantee that the sum is positive. However, we are always using a 
tiny fraction of states close to the singlet ground state and find that $\chi_C(T,N)$ converges from below with increasing 
$W_C(N)$. A satisfactory cutoff converges $\chi_C(T,N)$ to its peak. The $\chi_C(T,N)$ maxima in Fig.~\ref{fig2} and Fig.~\ref{fig3} 
are less converged than the $S_C(T,N)/T$ maxima.  

The spectrum in the $M = 0$ sector is the densest since it includes a Zeeman component of all states with $S > 0$, and it has the 
largest truncation error. The following results are mostly based on cutoffs $W_C(\alpha,N)$ that retain 10 states with $M = 5$ and 
none with $M > 5$. The $M = 0$ and 1 sectors contain more than 400 states,nearly 1000 states, when the Zeeman components include the projection from sectors 
with higher $M$ within cutoff $W_C(\alpha,N)$. The total number of states is $R_C = 4532$ and 2705 for $N = 48$ and 64, respectively at $\alpha = 2/3$, 
and 3647 and 2239 at 48 and 64 at $\alpha = 1/2$. The results are not sensitive to $W_C(N)$ provided the cutoff is high enough to enforce 
convergence at the maxima in Fig.~\ref{fig2} and Fig.~\ref{fig3}.

\subsection{\label{Sec:2C}Extrapolation}
\begin{figure}
\includegraphics[width=\columnwidth]{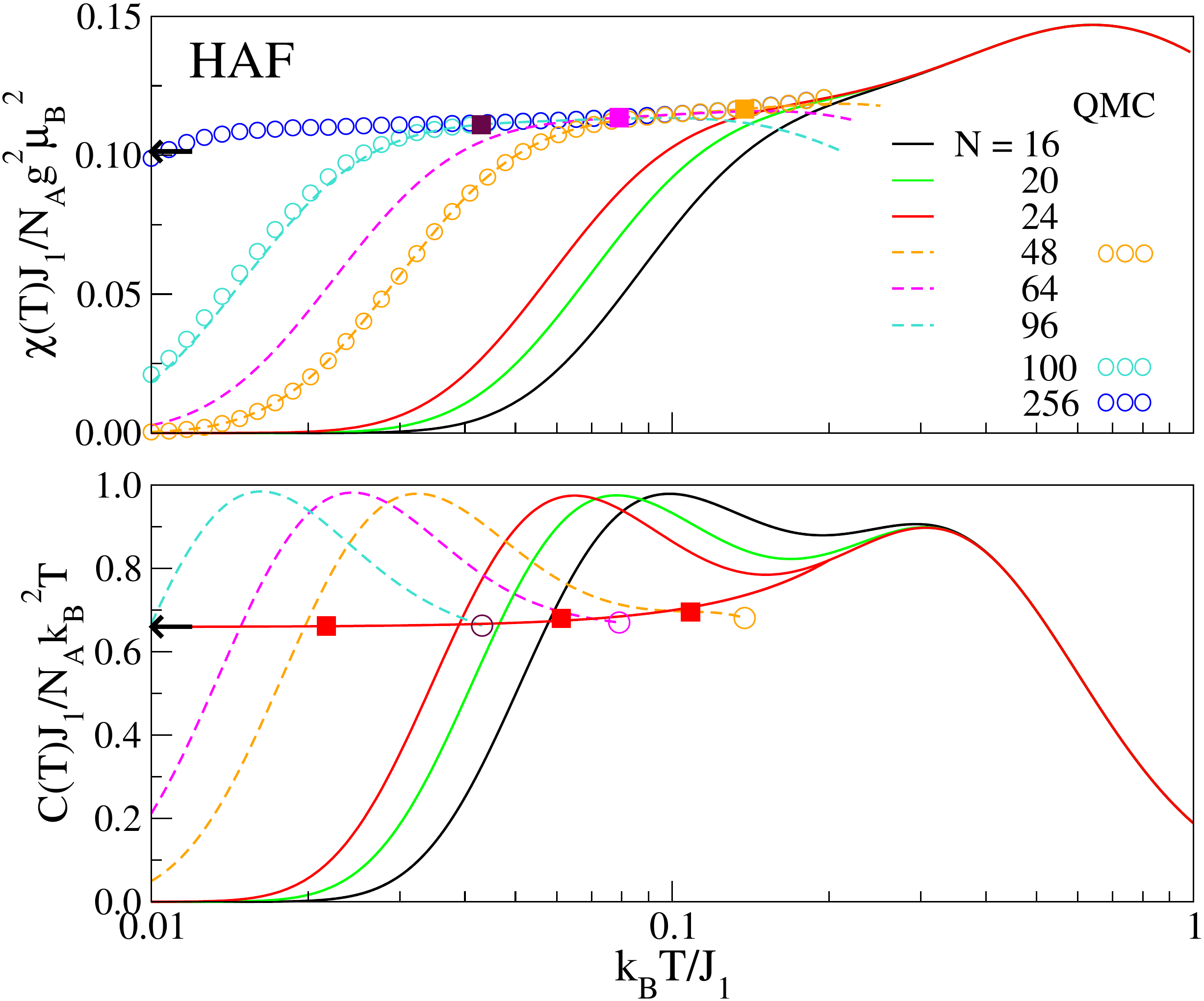}
\caption{\label{fig4} HAF results for $\chi(T)$ (upper panel) and $C(T)/T = S^\prime(T)$ (lower panel). 
Solid and dashes lines are ED and truncated DMRG, respectively, at the indicated system size $N$. The QMC 
calculations of $\chi(T)$ follow Ref.~\onlinecite{sandvik2010}. The arrows are exact at $T = 0$. The high $T$ peak is fully 
converged in both panels. The DMRG lines for $C(T,N)/T$ terminate at $T^\prime(N)$, the maximum of $S_C(T,N)/T$, 
shown as open circles (lower panel) and solid squares (upper panel). The squares and extrapolation in the 
lower panel are discussed in the text.}
\end{figure}

Fig.~\ref{fig4} shows the absolute spin susceptibility $\chi(T)$ and specific heat $C(T)$ of the HAF. $N_A$ is Avogadro's number, $\mu_B$ is the Bohr magneton and $g$ is the electronic $g$ factor. We use reduced units from here on and 
label the axes of subsequent graphs as $\chi(T)$ or $S^\prime(T) = C(T)/T$ vs. $T$. 

ED (solid lines) clearly indicates converged $\chi(T)$ at $T > T_n = 0.20$. The peak at $T_m = 0.641$ and $\chi(T_m) = 0.147$ in 
the upper panel are quantitative~\cite{johnston2000}. DMRG (dashed lines) extends $\chi(T)$ to lower $T$ and illustrates once 
again that finite-size gaps decrease with increasing system size. The squares on the DMRG curves are $\chi(T^\prime,N)$ evaluated at 
$T^\prime(N)$, the maximum of $S_C(T,N)/T$. Open symbols are quantum Monte Carlo (QCM) calculations following Ref.~\onlinecite{sandvik2010} 
at $N = 48$, 100 and 256. The arrow marks the exact $\chi(0) = 1/\pi^2$. There are logarithmic corrections~\cite{johnston2000} 
at $k_B T/J_1 < 0.005$.

The lower panel of Fig.~\ref{fig4} shows the entropy derivative, $S^\prime(T) = C(T)/T$, over the same range. The area under 
ED (solid) lines is $\ln2$ and ED again converges for $T > 0.20$. The peak at $T^\ast = 0.307$ and $S^\prime(T^\ast) = 0.897$ 
are quantitative~\cite{johnston2000}. The arrow marks the exact $S^\prime(0) = 2/3$. DMRG (dashed lines) terminate at $T^\prime(N)$, 
now shown as open circles. The $S^\prime(T,N)$ maxima are at $T_m(N) < T^\prime(N)$. We return later to the squares. DMRG and truncation is 
almost quantitative up to $T^\prime(N)$, as seen from ED at $N = 24$. That is also the case for $\chi(T)$ at $T^\prime(N)$ in the upper panel.

There are far fewer published $C(T,N)$ than $\chi(T,N)$ curves. Moreover, $C(T,N)$ plots completely obscure the behavior 
at low $T$ where finite size effects are responsible for \textit{deviations} from linearity. QMC works beautifully for $\chi(T)$ but 
produces scatter plots for $C(T,N)/T$ at low $T$; it is ill suited for narrow features such as the $T_m(N)$ peaks. Finite size 
effects are readily understood. Since $S(T_n)$ is in the thermodynamic limit and finite systems have $S^\prime(0,N) = 0$, reduced 
$S^\prime(T,N)$ at low $T$ must be compensated by increased $S^\prime(T,N)$ at $T < T_n$. The truncated $S^\prime(T,N)$ have maxima 
at $T_m(N)$ where
\begin{equation}
	S_C^\prime(T_m,N)=\frac{C_C(T_m,N)}{T_m} > S^\prime(T_m)
\label{eq:entropy_specific_trunc}
\end{equation}
Convergence of $S_C^\prime(T,N)$ to the thermodynamic limit is from above while $S_C(T,N)$ converges from below.

In order to extract the thermodynamic limit of $S^\prime(T)$, we note that its maximum $T^\ast$ is above $T_n$. 
The $S^\prime(T,N)$ peaks are superimposed on a smooth background that we take as $S^\prime(T) = a(1 + bT + c^2T^2)$ 
for $T \leq T_n$. There are three parameters, $a$, $b$ and $c$. Two are fixed by $S(T_n)$ and $S^\prime(T_n)$. The third is 
fixed by the scaling $x(N) = S^\prime(T_m)/S^\prime(T_m,N) < 1$ for each truncated spectrum. We sought parameters for which 
$x(N)$ is size independent. The best choice had $x$ between 68.5 and 69.4$\%$ for the $T_m(N)$ peaks from $N = 24$ to 96. 
The resulting $S^\prime(T) = C(T)/T$ is the $T < T_n$ line in Fig.~\ref{fig4}. We find $S^\prime(0) = 0.659$ and very small $b = 10^{-5}$. 
The exact result is $S^\prime(0) = 2/3$ and $S^\prime(T)$ is quadratic at low $T$~\cite{johnston2000} aside from logarithmic 
corrections below $T = 0.005$.

We conclude that hybrid ED/DMRG works well for the HAF’s spin susceptibility and specific heat. The HAF is especially simple: spin-1/2, 
one spin per unit cell, one exchange and hence no frustration. We did not appreciate that improved extrapolation is needed for the 
frustrated $J_1-J_2$ model in Eq.~\ref{eq:j1j2}. The $S^\prime(T,\alpha)/T$ peak at $T^\ast(\alpha)$ shifts to $T < T_n(\alpha)$ and 
reaches $T = 0$ near the critical point $\alpha_c$. Agreement with the HAF is necessary but not sufficient.

\section{\label{sec3} Thermodynamics of $J_1-J_2$ models}
In this section we study the $J_1-J_2$ model with $\alpha \geq \alpha_c = 1/4$ 
in Eq.~\ref{eq:j1j2}. Its quantum phases have already been mentioned. The general TMRG study of Lu et al.~\cite{xiang2006} has results for $J_1, J_2$
of either sign and discusses the thermodynamics of both singlet and FM phases. Sirker~\cite{sirker2010} later applied TMRG to the singlet 
phases of F/AF chains with $\alpha$ ranging from $\alpha_c$ to $2$. 
QMC is not applicable to frustrated interactions. The ground state is a singlet $(S = 0)$ and is doubly 
degenerate in the IC phase.  

ED up to 24 spins converges to the thermodynamic limit for $T > T_n(\alpha)$ as seen in Fig.~\ref{fig4} for the HAF. 
The $T_n(\alpha)$ in Table~\ref{tab3} are in units of $|J_1|/k_B$. They are based on $S^\prime(T,N) = C(T,N)/T$, whose size 
dependence is usually stronger than that of $\chi(T,N)$.  The increasing density of states in Fig.~\ref{fig1} with 
decreasing $\alpha$ accounts for an order of magnitude variation of $T_n(\alpha)$. The area per spin under $S^\prime(T,\alpha,N)$ 
curves is respectively $\ln 2$ for ED and $(\ln R_C(\alpha,N))/N$ for DMRG, where $R_C(\alpha,N)$ is the truncated number of 
states in Eq.~\ref{eq:num_state}.

\begin{table}
\caption{\label{tab3} Reduced temperature $T_n(\alpha)$ at which the thermodynamic limit of $S^\prime(T,\alpha) = C(T,\alpha)/T$
is reached for $N = 24$ spins in $J_1-J_2$ models with frustration $\alpha$ in Eq.~\ref{eq:j1j2}.}
\begin{ruledtabular}
\begin{tabular}{ c  c  c  c  }
Model, $\alpha$ & $T_n(\alpha)$ & $S(T_n,\alpha)$ & $S^\prime(T_n,\alpha)$  \\ \hline
$\alpha_c= 1/4$  &  0.02          &     0.413   & 2.665   \\
1/3              &  0.06          &     0.481   & 1.838   \\
1/2              &  0.14          &     0.399   & 1.656   \\
2/3              &  0.17          &     0.293   & 1.533   \\
HAF              &  0.20          &     0.143   & 0.820   \\
\end{tabular}
\end{ruledtabular}
\end{table}

The singlet quantum phases of spin-1/2 chains are either gapless with a nondegenerate ground state or gapped with a doubly 
degenerate ground state~\cite{allen1997}. The HAF is gapless while the $J_1-J_2$ model has both gapped and gapless singlet phases. 
The HAF has logarithmic contributions to $\chi(T)$ and $C(T)$ at $T < 5 \times 10^{-3}$ that are followed to several decades lower 
$T$ in Ref.~\onlinecite{johnston2000}. The gapped incommensurate (IC) phase runs from the exact quantum critical point~\cite{hamada88} 
$\alpha_c = 1/4$ to another critical point~\cite{soos2013, soos-jpcm-2016} around $\alpha = 0.806$. The IC gap $\Delta(\alpha)$ is 
exponentially small~\cite{itoi2001}, however, and has yet to be evaluated. The ground state degeneracy is followed numerically 
using DMRG with periodic boundary to compute the static structure factor $F(q,\alpha)$ at wave vector $q$~\cite{soos2013, soos-jpcm-2016}. 
The $F(q,\alpha)$ peaks at $\pm q(\alpha)$ shift in the IC phase from $q(1/4) = 0$ to $q(0.806) = \pi/2$. The decoupled 
phase~\cite{soos2013, soos-jpcm-2016} for $\alpha > 0.806$ is gapless and commensurate. Its singlet ground state is nondegenerate 
and has quasi-long-range order with $q = \pi/2$.

Neither logarithmic corrections nor an IC gap matters for the thermodynamics at $T > 0.01$. Returning to Table~\ref{tab3}, we note that 
the average value of $S^\prime(T_n,\alpha)$ up to $T_n(\alpha)$ is $S(T_n,\alpha)/T_n$ and does not depend on the actual form 
of $S^\prime(T_n,\alpha)$ in the interval. Since the average at $\alpha = 1/3$ is more than four times $S^\prime(T_n,1/3)$, we 
infer that $S^\prime(T,1/3)$ decreases with $T$. Although not as strongly, $S^\prime(T,1/2)$ and $S^\prime(T,2/3)$ also decrease 
with $T$ while the HAF has increasing $S^\prime(T)$ to $T^\ast > T_n$.

\subsection{\label{Sec:3A}Critical point, $\alpha_c = 1/4$}
\begin{figure}
\includegraphics[width=\columnwidth]{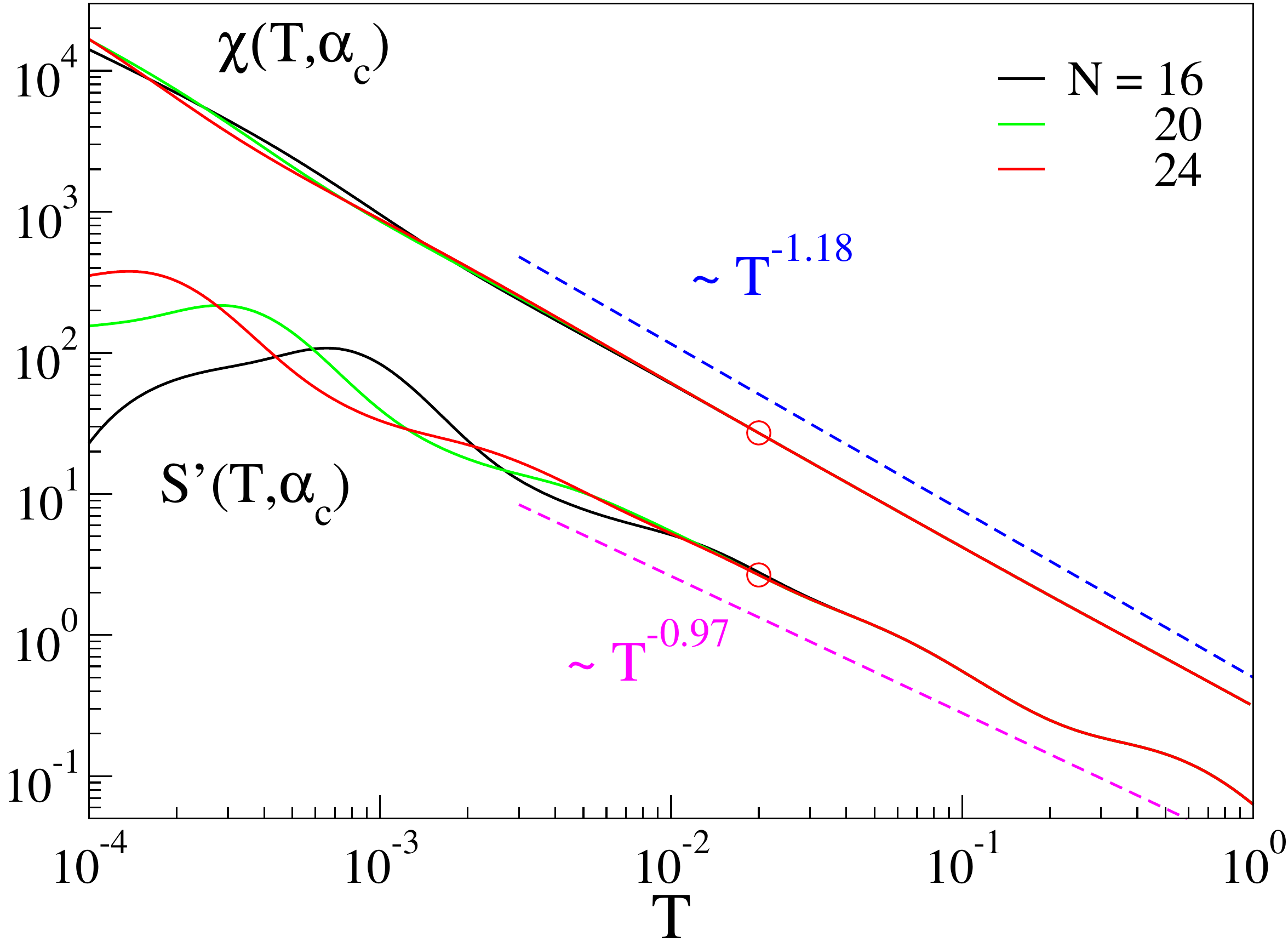}
\caption{\label{fig5} $\chi(T,\alpha_c)$ and $S^\prime(T,\alpha_c)$ in reduced units at the critical point, $\alpha_c = 1/4$. 
Open circles are at $T_n(\alpha_c) = 0.02$ where the thermodynamic limit is reached for ED up to 24 spins.}
\end{figure}

Thermodynamics at the critical point is remarkably different from larger $\alpha$. ED results in Fig.~\ref{fig5} for 
$S^\prime(T,\alpha_c)$ and $\chi(T,\alpha_c)$ in reduced units are almost power laws over several decades in $T$. 
The approximate exponents are $\gamma = -1.18$ and $-0.97$, respectively. ED to $N = 24$ at the critical point 
indicates that $T_n(\alpha_c) \sim 0.02$ and shows the stronger size dependence of $S^\prime(T,N)$. $S^\prime(T)$ is a 
measure of thermal fluctuations while $\chi(T)$ measures fluctuations of $M^2$, where $-S \leq M \leq S$ are the Zeeman 
levels of spin-$S$ states. 

Sirker et al.~\cite{sirker2011} studied the $J_1-J_2$ model in zero field on the FM side, $0 \leq \alpha \leq \alpha_c = 1/4$, 
using field theory and numerical methods. To leading order in $T$, the exact $\chi(T,\alpha_c)$ is $AT^{-4/3}$ with $A \sim 0.1685$ according to field 
theory and scaling for the classical model. Modified spin wave theory for the quantum model returns the same exponent with $A^\prime \sim 0.0992$. The reported susceptibility 
at $T = 10^{-3}$ is $\chi(10^{-3}) = A$ or $A^\prime \times 10^4$ while ED for for the
quantum chain of $N = 24$ gives $0.0882 \times 10^4$.
Such excellent agreement speaks to the accuracy of field theory and of ED at $N = 24$ for the thermodynamic limit at the critical point. On the 
other hand, the $\chi(T,\alpha_c)$ exponent in Fig.~\ref{fig5} deviates considerably from $-4/3$. TMRG~\cite{sirker2011} between $T = 0.003$ and 1 
gives slightly different exponents clustered around $T^{-1.2}$, consistent with Fig.~\ref{fig5}. Several reasons for the discrepancy were 
discussed~\cite{sirker2011}, including the possibility that TMRG could not reach sufficiently low $T$. The leading term of field theory is 
limited to $T < 10^{-3}$ and in this case ED to $N = 24$ reaches lower $T$.

To leading order in $T$, the field theory~\cite{sirker2011} free energy $A(\alpha_c)$ goes as $-T^{5/4}$. The entropy 
$S(T)$ and $S^\prime(T)$ go as $T^{1/4}$ and $T^{-3/4}$, respectively. The calculated 
exponent of $S^\prime(T,\alpha_c)$ is $\gamma \sim -0.97$ rather than $-0.75$ for $T > 0.01$. The $S^\prime(T,\alpha_c)$ exponent 
in this range is more negative than that of field theory while the  $\chi(T,\alpha_c)$ exponent is less negative. 
Hence $C(T,\alpha_c)$ goes as $T^{0.03}$ and is almost constant.

Field theory~\cite{sirker2011} indicates spectacular singularities at $\alpha_c$: $S^\prime(T,\alpha_c)$ and $\chi(T,\alpha_c)$ 
diverge at $T = 0$ while an IC gap $\Delta(\alpha)$ implies $S^\prime(0,\alpha) = \chi(0,\alpha) = 0$ for $\alpha > \alpha_c$. This is a mathematical result. In the 
present context, it is instructive to contrast $\alpha_c$ in Fig.~\ref{fig5} with the HAF in Fig.~\ref{fig4}. Increasing AF exchange $J_2$ 
over the range $\alpha_c < \alpha < \infty $ reduces both $\chi(T)$ and $C(T)/T$ by orders of magnitude at $T \sim 10^{-3}$ and 
by much less at $T \sim 0.1$, where thermal fluctuations are much stronger. The steep power-law $T$ decrease of both at $\alpha_c$ 
evolves into the weak $T$ dependence with a maxima at $T > 0$ in the HAF. Entropy conservation ensures the crossing of $S^\prime(T,\alpha)$ 
curves with different $\alpha$ while AF exchange accounts for $\chi(T,\alpha^\prime) < \chi(T,\alpha)$ when $\alpha^\prime > \alpha$. 
The qualitative changes from $\alpha_c$ to the HAF provide a framework for the thermodynamics at intermediate $\alpha$.

\subsection{\label{Sec:3B}Coupled sublattices, $\alpha = 2/3$}
\begin{figure}
\includegraphics[width=\columnwidth]{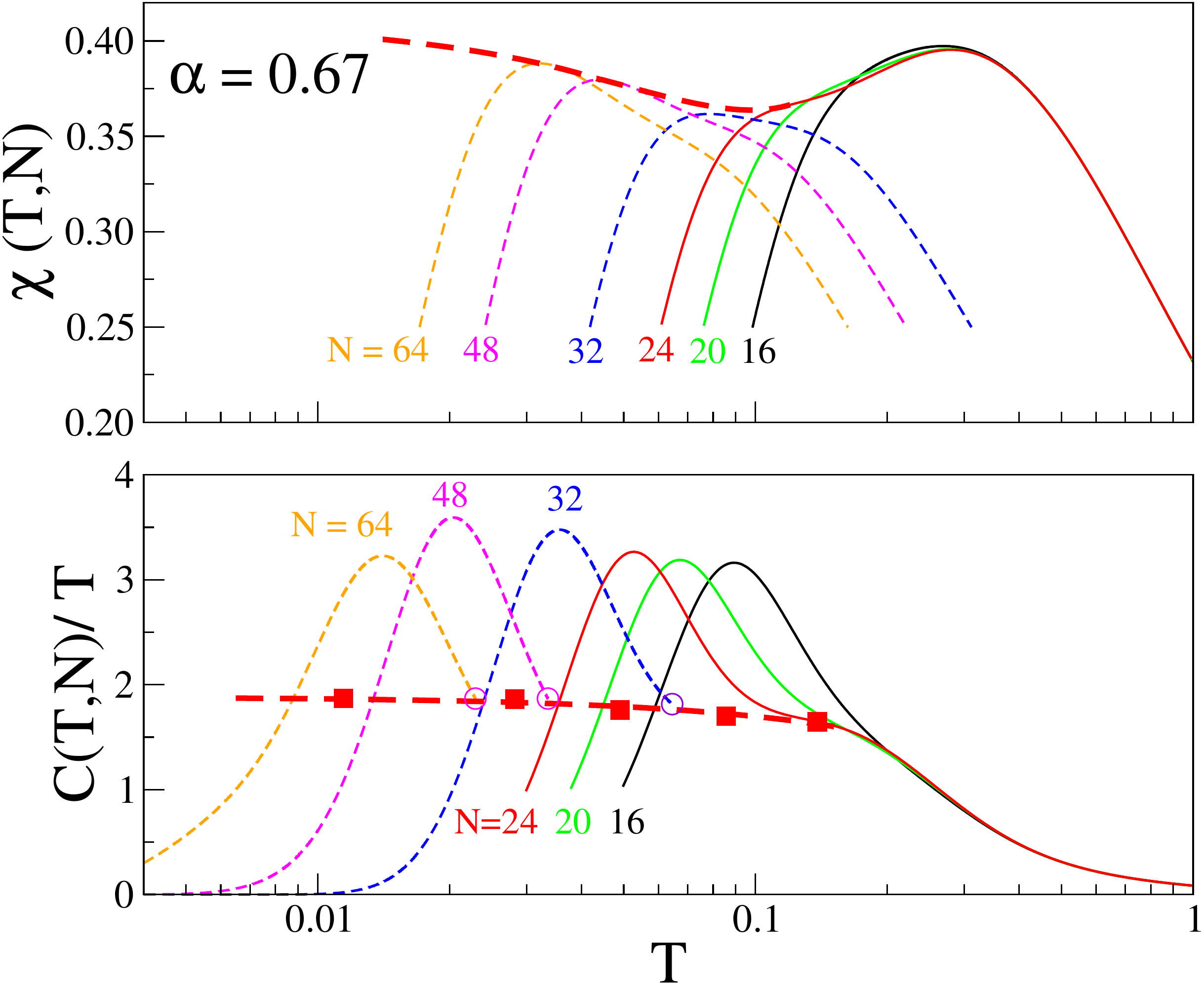}
\caption{\label{fig6} 
Upper panel: $\chi(T,N)$ for $\alpha= 2/3$ and $N$ spins in Eq.~\ref{eq:j1j2}. Solid and dashed lines are 
ED and DMRG. The bold dashed line is the estimated $\chi(T)$ in the thermodynamic limit. Lower panel: $C(T,N)/T$ 
for the same systems. Open circles are $T^\prime(N)$, the maxima in Eq.~\ref{eq:entropy_specific}. Squares 
are Eq.~\ref{eq:mvt}, the mean value in successive $T^\prime(N)$ intervals; the dashed line connecting 
them is to guide the eye.}
\end{figure}

The $\alpha \rightarrow \infty  (J_1 = 0)$ limit of Eq.~\ref{eq:j1j2} corresponds to HAFs on sublattices of odd and 
even numbered sites. Finite $J_1 < 0$ couples the HAFs and, as shown in Fig.~\ref{fig6} at $\alpha= 2/3$, increases 
both $\chi(T,\alpha)$ and $C(T,\alpha)/T$ compared to Fig.~\ref{fig4}. Finite size effects are more prominent and the 
HAF extrapolations no longer suffice. The reason is that $S(T,\alpha)/T$ either decreases monotonically or has a maximum 
at $T^\ast < T_n(\alpha,N)$. We consider an alternative analysis before discussing the $\alpha= 2/3$ results.

We suppress the model index $\alpha$ and recall that the truncated entropy $S_C(T,N)$ converges to $S(T,N)$ from below. 
The approximation that relates finite $N$ to the thermodynamic limit is
\begin{equation}
	\frac{S_C(T^\prime,N)}{T^\prime} \leq \frac{S(T^\prime,N)}{T^\prime} \leq \frac{S(T^\prime)}{T^\prime} 
	\equiv \langle S^\prime (T) \rangle_{T^\prime}
\label{eq:entropy_convrg}
\end{equation}
where $T^\prime(N)$ is the maximum defined in Eq.~\ref{eq:entropy_specific}. It follows that $T^\prime(N)$ is 
less than $T_n(N)$ but greater than $T_m(N)$, the maximum of $S^\prime(T,N)$ in 
Eq.~\ref{eq:entropy_specific_trunc}, where $S^\prime(T_m,N) > S^\prime(T_m)$. We note that $S_C(T^\prime,N)$ 
is a lower bound for $S(T^\prime)$ and use $T^\prime(N)$ to approximate the thermodynamic average $\langle S^\prime (T) \rangle$ 
between $T = 0$ and $T^\prime(N)$. Each system size generates a point at $T^\prime(N)$. It is convenient to define 
$T_1 = T^\prime(N_1)$ for the largest system, $T_2 = T^\prime(N_2)$ for the second largest, and so on. 

The mean value theorem can be applied to successive intervals to estimate
\begin{equation}
\begin{aligned}
	&  S^\prime(T_1/2) \approx \frac{S(T_1)}{T_1}, \qquad 0 \leq T \leq T_1  \\
	&  S^\prime(\left(T_1+T_2 \right)/2) \approx \frac {S(T_2)-S(T_1)} {T_2-T_1}, \quad T_1 \leq T \leq T_2 
\end{aligned}
\label{eq:mvt}
\end{equation}
and similarly at $T=\left(T_2+T_3 \right)/2$. This simple approximation is accurate when the size dependence of $S(T^\prime,N)/T^\prime$ is weak. 
The final point at $S(T_n)/T_n$ is in the thermodynamic limit, where $S^\prime(T_n)$ is also known. There is one 
input at each $T^\prime(N)$ and two at $T_n$ for estimating $S^\prime(T)$ up to $T_n$. The mean-value estimate could 
be replaced by linear, quadratic or other fits. That is premature, however, because Eq.~\ref{eq:entropy_convrg} 
returns an approximate $S(T^\prime)/T^\prime$ and experience with other models is needed first.

ED and DMRG results for $\chi(T,N)$ are shown in the upper panel of Fig.~\ref{fig6} for $H(2/3)$ in Eq.~\ref{eq:j1j2}. 
The thermodynamic limit holds for $T > T_n(2/3) = 0.17$. The $\chi(T)$ maximum at $T_m = 0.281$ is lower than 0.6413 for the HAF 
and $\chi(T_m) = 0.395$ is almost three times higher due to F exchange $J_1$. The bold dashed line that approximates the thermodynamic limit is linear extrapolation of 
the N = 48 and 64 maxima. The upturn of $\chi(T)$ at low $T$ is consistent 
with TMRG at $\alpha = 0.6$ in Fig.1 of Ref.~\onlinecite{sirker2010}. So are the magnitude at the peak and the lowest accessible $T$.

The lower panel of Fig.~\ref{fig6} shows $S^\prime(T) = C(T)/T$ and large finite-size peaks.
The DMRG curves stop at $T^\prime(N)$, the maximum of $S_C(T,2/3,N)/T$, which are shown as open circles. 
The squares are the mean value approximation, Eq.~\ref{eq:mvt}, which returns the squares in Fig.~\ref{fig4} (lower panel) 
when be applied to $S^\prime(T)$ for the HAF. We find $S^\prime(0) \sim 1.88$ at $\alpha= 2/3$, again about three times the 
HAF value. $S^\prime(T)$ gently decreases with $T$ at $\alpha= 2/3$ instead of gently increasing in the HAF.

\subsection{\label{Sec:3C}Incommensurate phase}
The $J_1-J_2$ model at $\alpha \geq 2/3$ can be viewed as HAFs on sublattices with F exchange $J_1 < 0$ reaching $-3J_2/2$ at $\alpha= 2/3$. 
The singlet ground state persists for more negative $J_1$ down to $\alpha_c= 1/4$ where as seen in Fig.~\ref{fig5} both $C(T)/T$ and $\chi(T)$ 
decrease sharply with increasing $T$. The $\alpha_c < \alpha< 2/3$ regime is particularly challenging. The thermodynamics is governed by 
weak AF exchange $J_2 < |J_1|$ at low $T$ and strong F exchange $J_1$ at high $T$.

\begin{figure}
\includegraphics[width=\columnwidth]{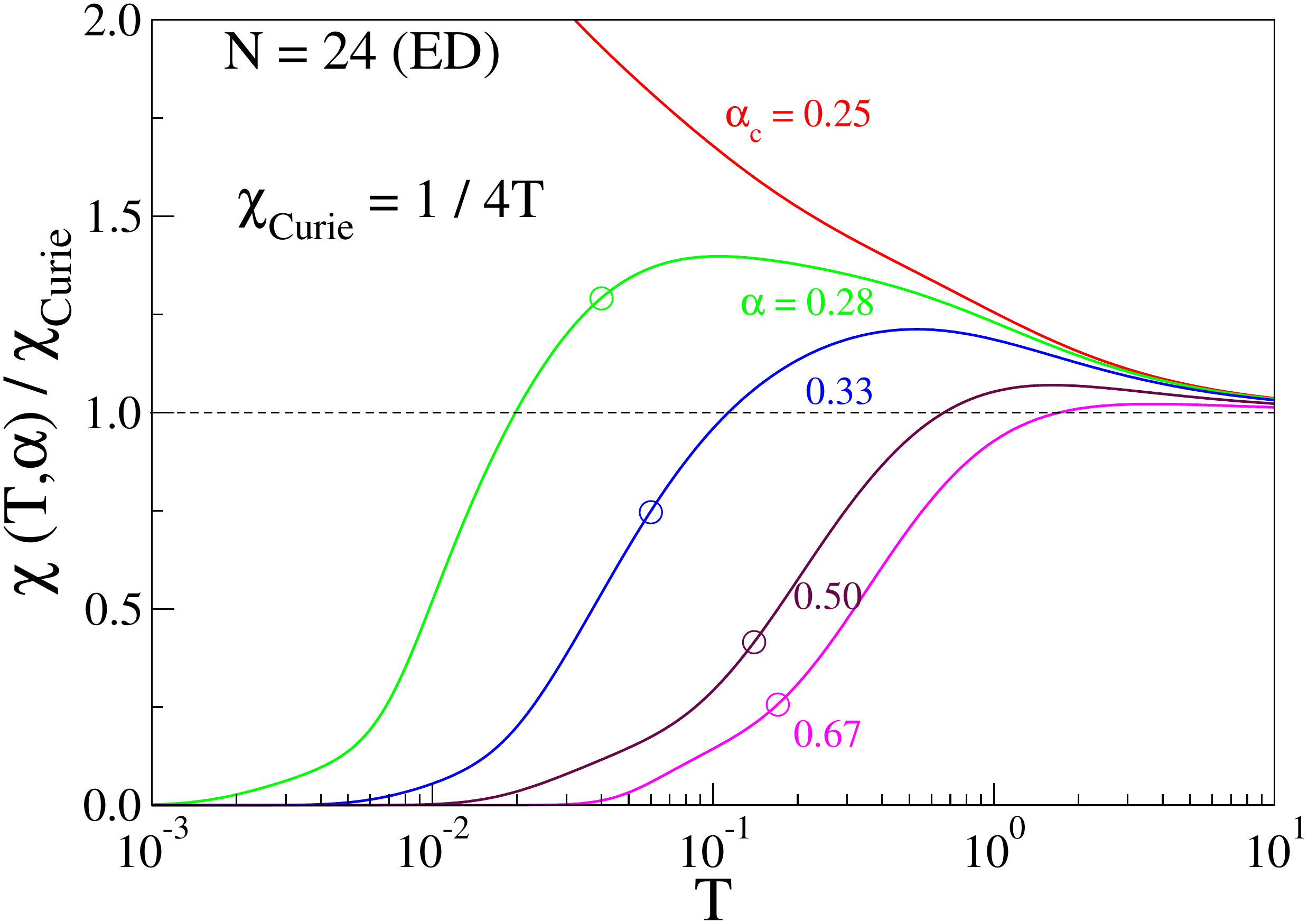}
\caption{\label{fig7} 
$4T\chi(T,\alpha)$ for 24 spins in Eq.~\ref{eq:j1j2}. Open circles mark $T_n(\alpha)$ where the thermodynamic limit is reached. 
The dashed line corresponds to free spins. Deviations above or below indicate that the net interaction is F or AF.}
\end{figure}

The Curie law for free spins is $\chi_C = 1/4T$ in reduced units. The $\chi(T,\alpha,24)/\chi_C$ curves in Fig.~\ref{fig7} 
deviate from free spins due to competing F and AF exchanges. The ``Curie temperatures" $T_C(\alpha)$ at which $\chi(T,\alpha)/\chi_C = 1$ are 
in the thermodynamic limit, above the $T_n(\alpha)$ in Table~\ref{tab3}. Offsetting F and AF exchanges lead to free-spin behavior at $T_C(\alpha)$, much 
as attractive and repulsive interactions in gases cancel at the Boyle temperature. The exact~\cite{sirker2011} $T^{-1/3}$ divergence at $\alpha_c$ is 
completely suppressed for $\alpha>\alpha_c$. The $T = 0$ limit of $\chi(T,\alpha)/\chi_C$ is zero for either gapless or gapped chains 
with singlet ground states.

Finite size gaps typically decrease roughly as $1/N$, but this expectation can fail in frustrated systems. The first (starred) excitation $E_2(N)$ 
in Table~\ref{tab2} for $\alpha= 1/2$ is twice as large at $N = 32$ than at $N = 24$. This singlet becomes degenerate with the ground state in the IC phase. 
The degeneracy for $N = 4n$ spins is limited to $n$ points $\alpha_j(N)$. The first is always $\alpha_c = 1/4$ while the last point $\alpha_n(N)$ increases 
with $N$. The $\alpha_j(N)$ are not distributed uniformly but are densest near the critical point~\cite{mk2012}. The gap $E_2(N,\alpha)$ at constant $\alpha$ 
varies randomly in large systems when $\alpha_n(N) > \alpha$. It vanishes when $\alpha=\alpha_j(N)$, is finite elsewhere, and decreases slowly with $N$ as 
the number of degenerate points increases. ED indicates~\cite{mk2012} that $\alpha_6(24) < 1/2 < \alpha_7(28)$ while DMRG shows~\cite{soos2013,soos-jpcm-2016} 
that $\alpha_{48}(192) = 0.66$. Hence $E_2(N)$ is already important at $N = 24$ for $\alpha = 1/2$ but not until much larger $N$ for $\alpha = 2/3$.

\begin{figure}
\includegraphics[width=\columnwidth]{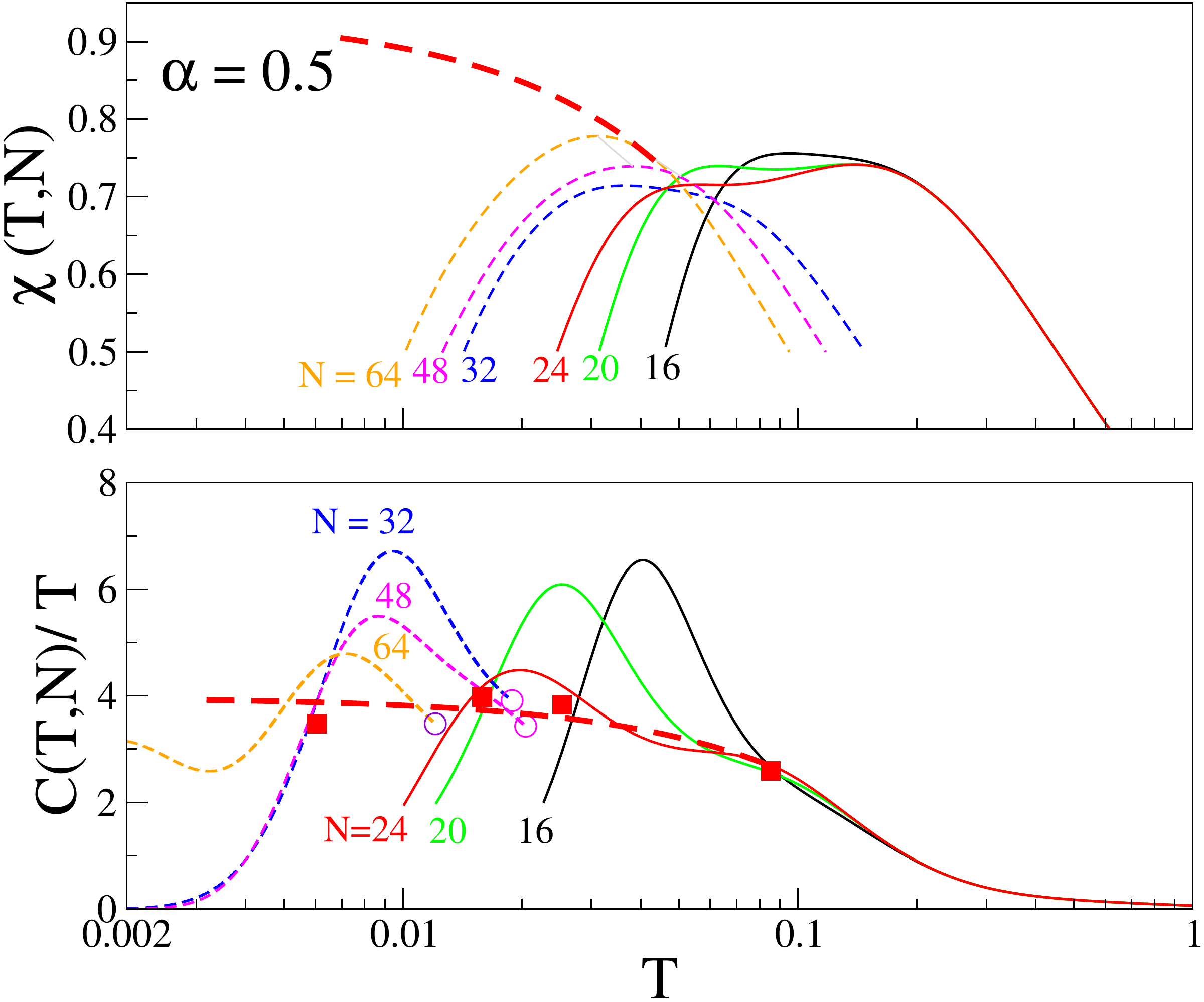}
\caption{\label{fig8}
Upper panel: $\chi(T,N)$ for $\alpha=1/2$ and $N$ spins in Eq.~\ref{eq:j1j2}. Solid and dashed lines are ED and DMRG. The bold dashed line is the
estimated $\chi(T)$ in the thermodynamic limit. Lower panel: $C(T,N)/T$ for the same systems. Open circles are $T^\prime(N)$, the maxima in
Eq.~\ref{eq:entropy_specific}. Squares are Eq.~\ref{eq:mvt}, the mean value in successive $T^\prime(N)$ intervals; the bold dashed line is to guide the eye.}
\end{figure}

Fig.~\ref{fig8} shows $\chi(T)$ and $C(T)/T$ curves at $\alpha= 1/2$. As expected, stronger F exchange compared to $\alpha= 2/3$ increases both and shifts 
them to lower $T$. The $\chi(T,N)$ peak increases with $N$ and shifts to lower $T$ at large $N$, but $\chi(T,N)$ decreases with $N$ at $T \sim 0.1$. The bold dashed line is linear extrapolation of the N = 48 and 64 peaks, shifted up slightly since since the thermodynamic limit is reached from below. It is quite approximate: $\chi(T,1/2) \sim 0.9$ at $T \sim 0$ and 
decreases to $\sim 0.7$ at $T \sim 0.06$. The weak maximum of 0.74 at $T = 0.14$ is in the thermodynamic limit. 

The $C(T)/T$ curves in the lower panel have similar $T^\prime(32) \sim T^\prime(48)$ that reflect the approximate nature of Eq.~\ref{eq:entropy_convrg}. 
We averaged both $T^\prime(32)$, $T^\prime(48)$ and $S(T^\prime,32)$, $S(T^\prime,48)$ to obtain the squares using Eq.~\ref{eq:mvt} 
for the mean values in the thermodynamic limit. The dashed line indicates linear $C(T) \sim 4T$ from $T=0.01$ to $0.03$ with downward deviation at
$0.04$ for $\alpha=0.5$ in the thermodynamic limit. TMRG~\cite{xiang2006,sirker2010} for $C(T)$ at $\alpha=0.4$ was extended~\cite{huang2012} 
down to $T=0.01$. As seen Fig.5 of Ref.~\onlinecite{huang2012}, 
$C(T) \sim 0.05$ at $T=0.01$. It is almost linear in $T$ up to $T=0.03$ and deviates downward at 0.04. The $T$ dependence is similar and $C(T)$ is known to increase at low $T$ 
with decreasing $\alpha$ in the singlet phase.

The degeneracies $\alpha_j(N)$ are closely spaced at $\alpha = 1/3$ and the excitations $E_j(N)$ are both small and dense. Numerical 
considerations discussed in Section~\ref{sec2} limit us to $N = 32$. On the other hand, the thermodynamic limit is already reached at 
$T_n(1/3) = 0.06$. 

\begin{figure}
\includegraphics[width=\columnwidth]{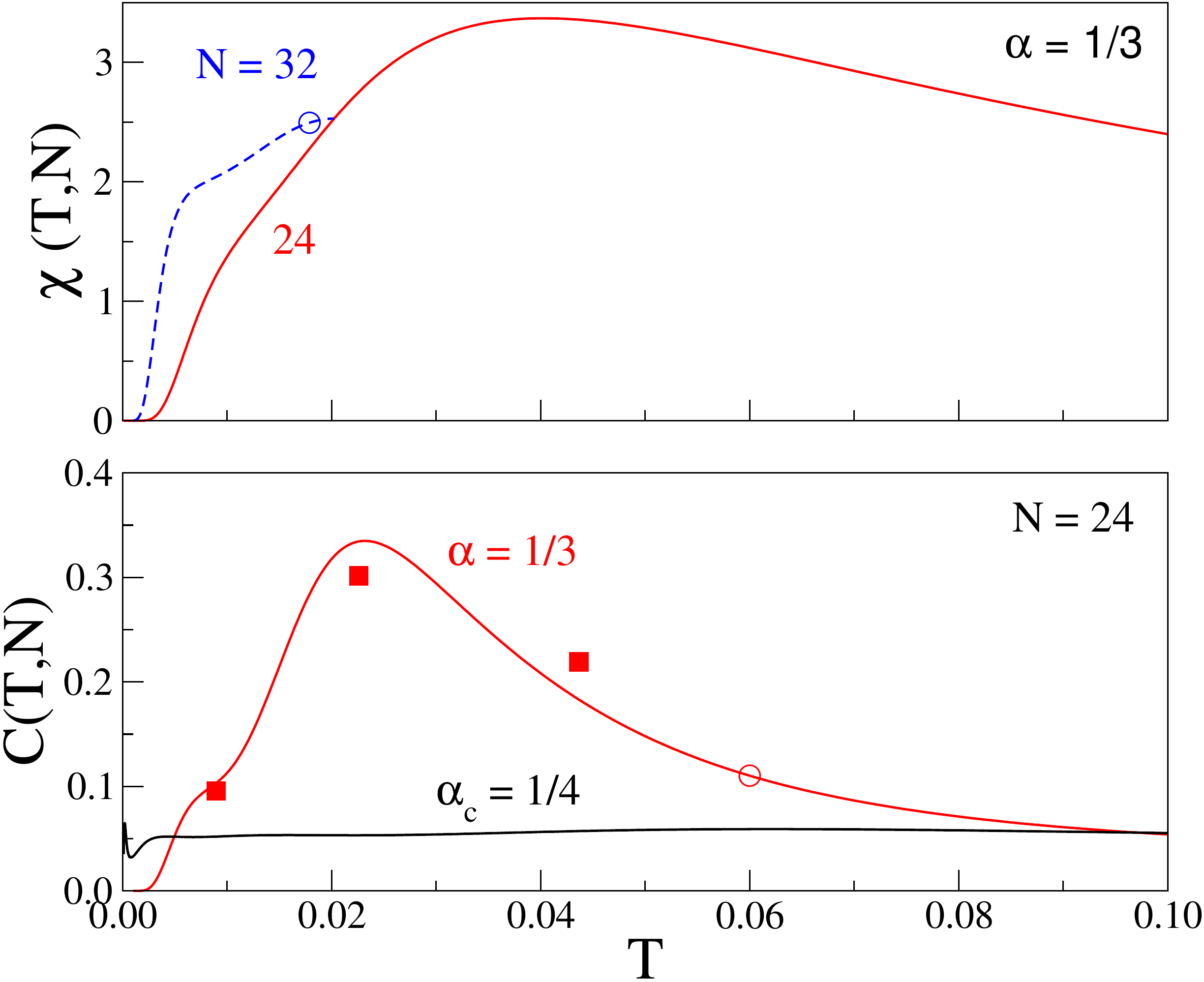}
\caption{\label{fig9}
Upper panel: $\chi(T,N)$ for $\alpha= 1/3$ in Eq.~\ref{eq:j1j2} and ED for $N = 24$ spins and DMRG for 32 spins up to the open circle at $T^\prime(32)$. 
Lower panel: $C(T,N)$ for $\alpha= 1/3$ and 1/4 for $N = 24$. The solid squares are based on DMRG for $N = 32$ and Eq.~\ref{eq:mvt} for three intervals based on 
$T^\prime(32)$,$T^\prime(24)$  and the open circle at $T_n(24) = 0.06$.}
\end{figure}

We switch in Fig.~\ref{fig9} to a linear $T$ scale for $\chi(T)$ and $C(T)$ up to $T = 0.10$, the Curie $T$ for free spins. 
AF correlations at lower $T$ lead to slower than $1/T$ increase of $\chi(T)$ and a maximum at $T \sim 0.04$. 
The truncated $\chi(T,32)$ peak confirms that $\chi(T)$ decreases in the thermodynamic limit at least to $T = 0.02$. The estimated $T \sim 0$ value of $\sim 1.7$ 
is more than 10 times that of the HAF. TMRG~\cite{sirker2010} at $\alpha= 0.3$ 
indicates a $\chi(T)$ maximum at $T \sim 0.02$. This is consistent with Fig.~\ref{fig9} since the peak shifts to $T = 0$ just above $\alpha_c = 1/4$.

We have three intervals for $C(T,N)/T$ based on $T^\prime(32) = 0.0179$, $T^\prime(24) = 0.0273$ and $T_n = 0.06$. The mean values using Eq.~\ref{eq:mvt} 
lead to the squares in the lower panel. The exact $C(T,1/3,24)$ calculation was first reported in Ref.~\onlinecite{meisner2006}. Our results for 
$T <T_n$ suggest that the peak in the thermodynamic limit is slightly lower and shifted to higher $T$. Also shown is the specific heat $C(T,1/4) \sim T^{0.03}$ 
at the critical point. It is almost constant in this interval since the $C(T)/T$ exponent in Fig.~\ref{fig5} is close to $-1$.

Spinless fermions~\cite{jordan1928} can be used to represent spin-1/2 chains; the $S^z = 0$ ground state corresponds to a half-filled band. The HAF 
has two-fermion interactions while the $J_1-J_2$ model, Eq.~\ref{eq:j1j2}, has up to four-fermion interactions. Both $C(T)/T = S^\prime(T)$ and $\chi(T)$ 
are proportional to the density of states at the Fermi energy as $T \rightarrow 0$. The Wilson-Sommerfeld ratio in reduced units is
\begin{equation}
	R_W(T)=\frac{4\pi^2 \chi(T)}{3S^\prime(T)}
\label{eq:wilson-sommerfeld}
\end{equation}
$R_W(T) = 1$ for free fermions, independent of $T$. The HAF result is $R_W(0) = 2$ with $10\%$ variations up to $T = 0.4$~\cite{johnston2000}. The $J_1-J_2$ model has 
increased $R_W(0.01) \sim 2.6$ at $\alpha= 2/3$ and $> 2.8$ at $\alpha= 1/2$. Much larger $R_W(0.01) \sim 150$ is found 
at $\alpha_c = 1/4$. $R_W(\sim0)$ increases when low-energy excitations have large $S$ because $C(T)/T$ contributions go as the $(2S + 1)$, the Zeeman degeneracy, 
while $\chi(T)$ contributions go as $S(S + 1)(2S +1)/3$, the sum over $M^2$.

\section{\label{sec4} Discussion}
We have presented a hybrid ED/DMRG approach to the thermodynamics of 1D models that never requires the full energy 
spectrum $\lbrace E(N) \rbrace$ of large systems and tested it in Section~\ref{sec2} against the spin-1/2 HAF. The $2^N$ 
states of spin-1/2 chains are found exactly in small systems and suffice for the thermodynamics at high $T$. DMRG for 
larger systems is used to obtain the lowest few thousand excitations $E_j(N)$. Thermodynamics at low $T$ is based on the 
truncated spectrum $E_j(N) \leq W_C(N)$. The cutoff criterion is convergence to the maximum of $S_C(T,N)/T$ and $\chi_C(T,N)$ 
with $T$, where $S_C(T,N)$ and $\chi_C(T,N)$ are respectively the truncated zero-field 
entropy and susceptibility per site. The 
thermodynamic limit at $T$ is approximated by maximum of $S_C(T,N)/T$ or of 
$\chi_C(T,N)$ at system size $N$. 

Exact diagonalization (ED) of the HAF with $N = 24$ spins becomes quantitative for $T \geq 0.20J/k_B$ as shown in Fig.~\ref{fig4}. 
DMRG up to $N = 96$ extends the thermodynamic limit for $\chi(T)$ and $S^\prime(T) = C(T)/T$ to an order of magnitude lower $T$, 
in excellent agreement with exact and numerical results. We are studying the performance of DMRG and truncation in 1D systems 
such as half-filled Hubbard, extended Hubbard and related models with charge as well as spin degrees of freedom. These models 
reduce to the HAF in the atomic limit. Charge degrees of freedom limit ED to smaller $N < 20$ with larger finite size gaps. There is 
greater scope for DMRG and truncation before running into the accuracy issues discussed in Section~\ref{Sec:2A}.

The motivation for this work was the thermodynamics of the frustrated $J_1-J_2$ model, Eq.~\ref{eq:j1j2},which is the starting point for the magnetic properties of several compounds with CuO$_2$ chains. 
F exchange $-J_1 > J_2$ is inferred~\cite{hase2004, drechsler2007, dutton2012, masuda2004, *park2007, wolter2012} at high $T$ from 
Curie-Weiss fits of $\chi(T)$ over a limited interval in which deviations from free spins in Fig.~\ref{fig7} are positive, but different $J_1$, 
$\alpha$ combinations return~\cite{xiang2006,dmsrz2018} similar $\chi(T)$. The net interaction is AF at low $T$ where an applied field can induce the FM 
state in some system. The $J_1-J_2$ model specifies the entire range of magnetization and magnetic specific heat. The data set~\cite{dutton2012} for LiCuSbO$_4$ were successfully modeled~\cite{dmsrz2018} by $|J_1| = –28.7$ K and 
$\alpha=2/3$ down to $T \sim 5$ K  $(T/|J_1| \sim 0.17)$ where finite-size gaps limit $N = 24$ results. Data below 5 K require improved 
thermodynamics as well as taking into account corrections to isotropic exchange and other magnetic interactions.

In an applied magnetic field, the $J_1-J_2$ model with anisotropic exchange supports a number of exotic quantum phases: 
IC, multipolar, vector chiral, among others~\cite{hikihara2008,sudan2009,aslam17}. The nature of the ground states, 
spin correlations and hidden symmetries are active areas of research, primarily of $T \rightarrow 0$ properties. 
That limit is beyond our approach. We alluded in the Introduction to mathematical and physical motivations. The CuO$_2$ 
chains have $-J_1 \sim 10^2$ K and anisotropic $g$-tensors that indicate $5-10\%$ deviations from isotropic exchange. 
Direct comparisons of the $J_1-J_2$ model, Eq.~\ref{eq:j1j2}, are limited to $T > 1$ K ($T > 0.01$ in reduced units), 
below which spin-orbit coupling and other magnetic interactions must be included. Considerably lower $T$ is relevant 
to exact field theory results at $\alpha_c$, for the gap $\Delta(\alpha)$ in the IC phase, or for logarithmic corrections. 
Quantitative analysis of magnetic data in the $1 - 10$ K range will be needed extract model parameters.

The hybrid ED/DMRG approach exploits the fact that the thermodynamic limit is reached at high $T$ in small 
systems that can be treated exactly. DMRG generates the excitations and truncated partition functions of 
increasingly large systems. We have focused on the spin susceptibility and specific heat of spin-1/2 chains. 
Other thermodynamic quantities are equally accessible, as indeed are applications to any 1D quantum cell model.\\
\begin{acknowledgments}
SKS thanks DST-INSPIRE for financial support.
MK thanks DST India for Ramanujan fellowship for financial support. 
\end{acknowledgments}
\end{document}